# The Extremal Index and the Maximum of a Dependent Stationary Pulse Load Process Observed above a High Threshold

Baidurya Bhattacharya[*]

*Department of Civil Engineering, Indian Institute of Technology ,Kharagpur, WB  721302, India*

**Abstract**

Observing a load process above high thresholds, modeling it as a pulse process with random occurrence times and magnitudes, and extrapolating life-time maximum or design loads from the data is a common task in structural reliability analyses.  In this paper, we consider a stationary live load sequence that arrive according to a dependent point process and allow for a weakened mixing-type dependence in the load pulse magnitudes that asymptotically decreases to zero with increasing separation in the sequence.  Inclusion of dependence in the model eliminates the unnecessary conservatism introduced by the i.i.d. (independent and identically distributed) assumption often made in determining maximum live load distribution.  The scale of fluctuation of the loading process is used to identify clusters of exceedances above high thresholds which in turn is used to estimate the extremal index of the process.  A Bayesian updating of the empirical distribution function, derived from the distribution of order statistics in a dependent stationary series, is performed.  The pulse arrival instants are modeled as a Cox process goverened by a stationary lognormal intensity.  An illustrative example utilizes in-service peak strain data from ambient traffic collected on a high volume highway bridge, and analyzes the asymptotic behavior of the maximum load.

---

[*] Corresponding author.  Email: baidurya@iitkgp.ac.in, Fax: +91 3222 282254

# 1. Introduction

A general formulation of the time-dependent reliability function of a structural component at time *t,* Rel(*t*), under an "overload" type limit state can be given by:

$$\text{Rel}(t) = P[R(\tau) - D - L(\tau) - W(\tau) - S(\tau) - ... > 0 \text{ for all } \tau \in [0,t]] \tag{1}$$

*R* is the resistance of the structural component, possibly time dependent due to aging effects. *L, W, S* etc. represent time-varying live, wind, snow loads and so on, respectively. The dead load (*D*) is generally assumed not to vary with time. Evaluation of the first passage probability in Eq. (1) is involved but can be simplified by considering some realistic and practical load combination scheme [1-3]. Further simplification is possible if $R(\tau)$ is replaced by a representative resistance, $R_e$, that is independent of time. In almost every relevant load case, the life-time maximum of one of these load processes must be evaluated [4]. A typical example can be formulated as:

$$R_e - D - L_{\max,t} = 0 \tag{2}$$

where $L_{\max,t}$ is the maximum live load on the structure during (0,*t*]. Live loads typically have discontinuous sample functions with jumps occurring at random instants of time, $\tau_i$, with random magnitudes, $L_i$. If the duration of the individual live load is small compared to the life of the component (or if a filtering with a high threshold is employed in data collection), the live load process may be idealized as a pulse process. In such idealized situations, the maximum of the observed live load magnitudes,

$$L_{\max,t} = \max\{L_1, L_2, \ldots, L_{N_t}\} \tag{3}$$

is of importance. $N_t$ is the random number of loading events during the interval $[0,t]$. Similar formulations can be made for maximum wind, wave or earthquake loads, transient pressure/temperature peaks in a loss of coolant accident etc.

Since the number of events in any bounded time interval is finite, the loading events over a time interval $[0, t]$ can be described as a marked ordinary point process, $N(t)$, where the marks are the live load magnitudes: $L_1, L_2, ..., L_{N_t}$. Let the random occurrence times of the point process be $\tau_1, \tau_2, ..., \tau_{N_t}$. The point process is ordinary (i.e., not more than one event occur in an infinitesimal interval of time) due to the finite response time of the recording device. It should be noted that the marks are random in nature, and the number of events, $N_t$, over the interval $[0, t]$ is a random variable as well.

The simplest model for predicting the life-time maximum distribution from an observed sequence $L_1, L_2, ..., L_n$ ($n$ finite) is to assume that the marks, $L_i$, are mutually statistically independent and are identically distributed (the so-called "i.i.d." assumption) and independent of the occurrence times as well. These assumptions (i) facilitate the estimation of the parent distribution function, $F$, of $L_i$, (ii) allows the modeling of threshold exceedances as a Bernoulli sequence, and (iii) yields "return period" values as well as the distribution function of the maxima corresponding to a known finite number of life-time occurrences. In many practical instances, each member of the observed sequence $L_1, L_2, ..., L_n$ actually represents the "block maximum" (e.g., annual maximum wind speeds or three-hourly maximum wave heights etc.) to facilitate computations; the cost, however, is the wastage of a large number of potentially useful data points.

Powerful generalizations can be made when the sample size, $n$, of the i.i.d. sequence approaches infinity. Under very general conditions that are satisfied by most parent distributions, the maximum, $L_{\max,t}$, of the i.i.d. sequence approaches one of the three classical extreme value distributions. Further, exceedances of the sequence $\{L_n\}$ above high thresholds $\{u_n\}$ approach the generalized Pareto distribution. Also, even if the occurrence times did not originally constitute a renewal process, they do become rarer on the time axis and approach the Poisson distribution as the said threshold becomes higher. The "peaks-over-threshold" method, based on the above assumptions, has been successfully applied in modeling extreme loads [5-7].

Nevertheless, the i.i.d. assumption cannot be always justified in maximum live load modeling (as has been warned recently by [8]). There may be significant dependence in the loading sequence (both in regard to load magnitudes and occurrence times) that would make the results from classical i.i.d. analyses conservative at best and erroneous in general. This paper builds up on the rich collection of results arising from the i.i.d. case, and presents a methodology that accounts for more generalized loading sequences including dependence in the occurrence point process and its associated marks. The formulation will be restricted to stationary and aperiodic sequences. The methodology will be demonstrated by predicting maximum live load effects on a highway bridge based on strain response data collected under ambient traffic.

## 2. Time-dependent Maximum Live Load Distribution from a Stationary Dependent Sequence

We now formalize the assumptions and restrictions regarding the strictly stationary but dependent sequence, $\{L_n\}$ with common marginal distribution $F$. Let $M_n$ denote the maximum of the sequence $\{L_n\}$. The dependence structure of $\{L_n\}$ is such that Leadbetter's [9] Conditions $D(u_n)$ and $D'(u_n)$ are satisfied.

Condition $D(u_n)$ is a type of distributional mixing (a much weakened form of strong mixing) and ensures that subsequences of $\{L_n\}$ becomes asymptotically independent with increasing separation between them. Let $F_{i_1,\cdots,i_n}(x)$ denote the joint distribution function of $L_{i_1},\ldots,L_{i_n}$ evaluated at the common point $x$. Condition $D(u_n)$ is said to hold for some given real sequence $\{u_n\}$, if for any integers $1 \leq i_1 < \cdots < i_p < j_1 < \cdots < j_{p'} \leq n$ for which $j_1 - i_p \geq l$, we have

$$\left| F_{i_1,\cdots,i_p,j_1,\cdots,j_{p'}}(u_n) - F_{i_1,\cdots,i_p}(u_n) F_{j_1,\cdots,j_{p'}}(u_n) \right| \leq \alpha_{n,l} \tag{4}$$

and $\alpha_{n,l_n} \to 0$ as $n \to \infty$ for some sequence $l_n = o(n)$.

The second of the two conditions, Condition $D'(u_n)$, limits the possibility of clustering of exceedances from the sequence above a high threshold. The Condition $D'(u_n)$ is said to hold for a strictly stationary sequence $\{L_n\}$ if for some given real sequence $\{u_n\}$,

$$\limsup_{n \to 0} n \sum_{j=2}^{[n/k]} P\{L_1 > u_n, L_j > u_n\} = 0 \text{ as } k \to \infty \tag{5}$$

where [ ] denotes the integer part.

The importance of Conditions $D(u_n)$ and $D'(u_n)$ is such that, under them, the dependent and stationary sequence $\{L_n\}$ has the following two very appealing properties [9]. Further, the strength of the dependence can be explicitly measured through a single parameter – the extremal index of the process.

(i) The asymptotic distribution of $M_n$, the maximum of $\{L_n\}$, has only three possible forms that are identical to the three classical extreme value distribution types, $H_c$, elicited below (the *classical* case arises from a sequence of i.i.d. random variables). The rate of convergence, however, is slower than in the i.i.d. case, and can be quantified using the *extremal index*, $\theta$, of the sequence, defined below.

(ii) The point process constituting the instants when the sequence $\{L_n\}$ exceeds the threshold $u_n$ converges in distribution to a Poisson process as *n* increases. Consequent of the asymptotic Poisson behavior, the maxima in disjoint time intervals become asymptotically independent as well.

*2.1 The extremal index*

The extremal index of a sequence can be interpreted as the reciprocal of the mean limiting cluster size above high thresholds, and is given by [9]:

$$\lim_{n\to\infty} P\{M_n \leq u_n(\tau)\} = \exp(-\theta\tau) \ , \ \tau > 0, \ 0 \leq \theta \leq 1 \tag{6}$$

if for some $\tau > 0$ there exists sequence $\{u_n(\tau)\}$ such that

$$\tau = \lim_{n\to\infty} n\left(1 - F(u_n(\tau))\right) \tag{7}$$

The extremal index, which is a number between 0 and 1, measures the strength of the dependence in the sequence $\{L_n\}$. Heuristically, $\theta = 0$ corresponds to an infinitely long memory sequence, $0 <$ $\theta < 1$ corresponds to a short memory sequence, and $\theta = 1$ corresponds to a memoryless sequence [10].

It is convenient at this point to introduce the associated sequence $\{\hat{L}_n\}$ that is i.i.d. and has the same marginal (or "parent") distribution, $F$, as the original sequence $\{L_n\}$. Let $\hat{M}_n$ be the maximum of the i.i.d. sequence $\{\hat{L}_n\}$. Classical extreme value theory (see e.g.,[11, 12]) gives the well-known result that under suitable regularity conditions that are satisfied by most common parent distributions, $F$, the distribution of the maximum $\hat{M}_n$, when suitably normalized by constants $\{a_n\}$ and $\{b_n\}$, converges to one of the three classical types, $H_c$:

$$P\left[\hat{M}_n \leq a_n x + b_n\right] \to H_c(z) = \exp\left[-(1+cz)^{-1/c}\right], \quad 1+cz > 0 \qquad (8)$$

where $z = (x-\varepsilon)/\delta$ in which $\varepsilon$ and $\delta > 0$ are appropriate location and scale parameters of the distribution. Two distribution $G$ and $H$ are of the same *type* if there are constants $a$ and $b > 0$ such that $G(x) = H(a+bx)$. Eq. (8) represents the generalized extreme value distribution, in which the parameter $c$ determines the type of the distribution: It is of (i) Type I (the Gumbel type) if $c = 0$, where $H_c$ is interpreted as the limit $\exp(-\exp(-z))$ as $c \to 0$, (ii) Type II (the Frechet type) if $c > 0$, and (iii) Type III (the Weibull type) if $c < 0$.

If Eq. (8) holds, the distribution of $M_n$ also converges, with the same set of normalizing constants $\{a_n\}$ and $\{b_n\}$ (or with one set altered) as above, to the type of $H_c^\theta$, where the exponent $\theta > 0$ is the extremal index of $\{L_n\}$:

$$P[M_n \leq a_n x + b_n] \to H_c^\theta(z) \tag{9}$$

The converse is also true. Clearly, $H_c$ and $H_c^\theta$ are of the same type for any given value of *c*. The significance of this result is that the distribution of the maximum $M_n$ of a stationary dependent sequence, provided it converges (which can be guaranteed by Conditions $D(u_n)$ and $D'(u_n)$), may be estimated, at least in the right tail, simply with the help of the marginal distribution *F* and the extremal index $\theta$ of the underlying process, as:

$$P[M_n \leq u_n] \approx F^{n\theta}(u_n) \tag{10}$$

for sufficiently high $u_n$ and large but finite *n*.

Eq. (10) is significant also as it highlights the degree of conservatism that may be introduced by the common and sometimes indiscriminate engineering practice of assuming a sequence to be i.i.d. when estimating the distribution of its maximum. If a sequence is i.i.d., its extremal index is 1 (the converse unfortunately is not necessarily true), and the distribution of the maximum is exactly $F^n$. For any other value of $\theta$ in the range (0,1),

$$P[M_n \leq u_n] > P[\hat{M}_n \leq u_n] \tag{11}$$

for every $u_n$ for which $0 < F(u_n) < 1$; and in the limit:

$$H_c^\theta(a+bz) > H_c(a+bz), \quad 0 < \theta < 1, \ 0 < H_c < 1 \tag{12}$$

for any constants *a* and *b*. In other words, the distribution of the maximum under the i.i.d. assumption is always to the right of that of the actual maximum for stationary dependent sequences. Since we are concerned with the life-time maximum live load, an erroneous adoption of the i.i.d. hypothesis would simply overestimate the demand; the degree of this conservatism depends solely on the value of the extremal index.

It is therefore clear that, from a practical point of view, the distribution of the maximum of a stationary dependent sequence may conveniently be estimated in two steps: first by obtaining the distribution of the maximum of the associated i.i.d. sequence, and second, by estimating the extremal index of the parent process. A practical method for estimating the extremal index is discussed next. The estimation of the associated i.i.d. sequence will be taken up subsequently.

There are two acceptable methods for estimating $\theta$ : the *blocks* method and the *runs* method. Under suitable conditions, both estimators are consistent for the extremal index, but we choose the runs method because it usually has the smaller bias of the two [13, 14]. Considering the extremal index as threshold dependent and estimating it at various values of the threshold has also been suggested [14, 15], but we do not adopt that approach in this paper.

Defining $M_{p,q} = \max\{L_p, \ldots, L_q\}$, the runs estimator of the extremal index, $\hat{\theta}_R$, is given by:

$$\hat{\theta}_R(x;r,n) = \frac{\sum_{i=1}^{n} I_{B,i}(L_i > x \geq M_{i+1,i+r-1})}{\sum_{i=1}^{n} I_{A,i}(L_i > x)} \quad , \quad r \geq 2 \tag{13}$$

in which $I_{A,i}(\cdot)$ and $I_{B,i}(\cdot)$ are indicator functions verifying the truth of the respective condition in parentheses. The estimate is basically the reciprocal of the average cluster size above high thresholds ($x$) in which two consecutive exceedances are part of the same cluster if they are less than $r$ observations apart (i.e., a *run* of observations below $x$ of length $r$ or greater are deemed to separate two adjacent clusters). Eq. (13) is based on the property of dependent stationary sequences that, for some appropriately chosen $r$ [10],

$$P\left[M_{2,r} \leq x | L_1 > x\right] = \theta + R(q(x)) \qquad (14)$$

such that

$$R(q(x)) \to 0 \qquad (15)$$

as $q(x) = 1 - F(x) \to 0$ (i.e., as $x \to x_0$ where $x_0 = \sup\{x : F(x) < 1\}$).

The quality of the estimate in Eq. (13) depends on $R$ which, however, is not known a priori. Nevertheless, it has been suggested [10] that $R$ has a simple log-linear form for a large class of processes:

$$R(q) \sim \beta_1 q^{\beta_2}, \quad \beta_1 \neq 0, \beta_2 > 0, \text{ as } q \to 0 \qquad (16)$$

This relation, in conjunction with the set of estimates obtained using Eq. (13) for a range of $x$, may be used to obtain a minimum squared error estimate for $\theta$ in Eq. (14).

Note that $\hat{\theta}_R$ depends also on the run length, $r$, a parameter that must be chosen with care. Attempts have been made to provide optimal estimates of $r$ based on minimum absolute bias considerations [14]. $r$ must be short enough so as not to group relatively independent observations in the same cluster, at the same time long enough to reflect the dependence structure of the underlying physical nature of the sequence, e.g., the average storm length in case of wave data. We propose Vanmarcke's [16] scale of fluctuation, $\tau_c$, as an estimate of the run length, $r$:

$$\hat{r} \approx \tau_c = \lim_{T \to \infty} T \gamma(T) \qquad (17)$$

where $\gamma(T)$ is the variance function of a stationary process $X(t)$ and is defined as the ratio of the variance of the local average over a window of length $T$ and the variance, $\sigma^2$, of the process $X(t)$:

$$\gamma(T) = \frac{\text{var}\left[(1/T)\int_{t-T/2}^{t+T/2} X(u)du\right]}{\sigma^2} = \frac{2}{T}\int_0^T \left(1 - \frac{\tau}{T}\right)\rho(\tau)d\tau \qquad (18)$$

where $\rho(\tau)$ is the autocorrelation function of $X(t)$. As noted in [16], $T/\tau_c$ may be interpreted as the "equivalent number of independent observations" contained in a sampling interval $T$; hence Eq. (17) is a valid estimate for $r$ because the run length essentially reduces the number of peaks in a dependent sequence to an equivalent number of asymptotically independent clusters. We therefore use the following integration to estimate the run length:

$$\hat{r} \approx 2\int_0^T \left(1 - \frac{\tau}{T}\right)\rho(\tau)d\tau, \quad r \geq 2, \text{ (}r\text{ integer)} \qquad (19)$$

for large *T*.

*2.2 Modeling the random load occurrence times*

The asymptotic behavior of the maximum of the marks of the point process provides the first part to the solution to the time-dependent maximum live load distribution problem. To complete the solution, the statistical description of the number of occurrences during the given time interval is required.

The simplest model for point processes – the Poisson model i.e., a renewal process with exponential inter-arrival times or, equivalently, a process with independent increments and a constant rate of occurrence – is analytically attractive, but may prove too simplistic for most loading processes. Nevertheless, the appeal of the pure Poisson process lies in its derivatives: it can be used as the building block for a large variety of processes showing clustering, dependence, non-stationarity etc. Clustering phenomena can be accounted for by the Neymann-Scott and the Bartlett-Lewis processes [17]. In the former, the points of a pure Poisson process act as cluster centers so that a random number of cluster points are distributed independently and identically around each cluster center. In the Bartlett-Lewis process, on the other hand, the cluster points are generated according to a finite renewal process around the original Poisson process. A Polya process, which is a non-stationary version of the pure birth process, can also be used to model clustering [1].

A more versatile generalization of the pure Poisson process occurs if the rate, $\Lambda(t) \geq 0$, itself is considered to be a random process yielding what is known as a doubly stochastic Poisson process (or Cox process) [17]. The mean measure of the point process in the interval [0,*t*] is a random variable and is given by:

$$M_t = \int_0^t \Lambda(\tau)d\tau \tag{20}$$

Then, conditioned on $M_t = m_t$ (where $m_t$ is any positive integer for given $t$), the point process $N(t)$ becomes a (generally non-homogeneous) Poisson process, i.e., the counts are distributed according to:

$$P[N(t) = x \mid M_t = m_t] = e^{-m_t} \frac{(m_t)^x}{x!} \tag{21}$$

Special cases of the Cox process include the following: $\Lambda(t) = \lambda$ (a constant) yield the *homogeneous* Poisson process; if $\Lambda(t) = \lambda(t)$ is a non-random function of time, we get a *non-homogeneous* Poisson process; and if $\Lambda(t) = \Lambda$ is a time-independent random variable, we are left with a *mixed* Poisson process.

We choose the Cox process to model the stochastic arrival rate for the load pulses first for its versatility, and then due to the asymptotic Poisson nature of the filtered point process above high thresholds that is consistent with the conditional Poisson characteristic of the Cox process. There is a rich collection of results pertaining to modeling the random rate function, $\Lambda(t)$; the two most common of which are the Markov modulated Poisson process [18] and the exponentiated Gaussian process (i.e. the lognormal process) [19].

We can now look at the maximum, $\hat{L}_{\max,t}$, of the associated i.i.d. peak strain sequence $\hat{L}_1, \hat{L}_2, ..., \hat{L}_{N_t}$. For fixed $N_t = n$ and $M_t = m_t$, the distribution of $\hat{L}_{\max,t}$ is,

$$P\left[\hat{L}_{\max,t} \leq x \mid N_t = n, M_t = m_t\right] = (F(x))^n \tag{22}$$

Removing the conditioning on $N_t = n$ first, we obtain:

$$P\left[\hat{L}_{\max,t} \leq x \mid M_t = m_t\right] = \exp\left[-m_t \{1 - F(x)\}\right] \tag{23}$$

Further, if the distribution of the mean measure, $M_t$, is known, the unconditional distribution of $\hat{L}_{\max,t}$ can be given by:

$$P\left[\hat{L}_{\max,t} \leq x\right] = \int_0^\infty \exp\left[-m_t \{1 - F(x)\}\right] f_{M_t}(m_t)\, dm_t \tag{24}$$

*2.3 Inclusion of sampling-related uncertainties*

The maximum live load is estimated based on observed data. Hence, it is important that uncertainties due to sampling be accounted for. For any given $l$, the true value of $F$ is unknown (see, e.g., [20]), hence we can describe it as a random variable $P$ with (prior) probability density function $f_P'$. The unknown $P$ is estimated from the sample as:

$$\hat{p}(l) \equiv \hat{F}(l) = \frac{1}{n+1} \sum_{k=1}^n \mathbf{I}(L_k \leq l) \tag{25}$$

Since the $L_k$'s are stationary, $\hat{p}$ is an asymptotically unbiased estimator of $F$ regardless of the fact that the $L_k$'s form a dependent sequence, although its variance is larger than that in the i.i.d. case.

Based on the $n$ observations, $\underline{I}$, of the indicator function $\mathbf{I}$, we can perform a Bayesian updating of the probability law of $P$ and obtain its posterior (updated) density function $f_P^{''}(p) = f_{P|\mathbf{I}=\underline{I}}(p)$ as:

$$f_P^{''}(p) = \frac{1}{C} \mathsf{L}(p;\underline{I}) f_P^{'}(p) \tag{26}$$

where $C$ is the normalizing constant, $\mathsf{L}(p;\underline{I}) = P[\mathbf{I}=\underline{I}|P=p]$ is the likelihood function and $f_P^{'}(p)$ is the prior probability density function of $P$. $\mathsf{L}(p;\underline{I})$ can be interpreted as the probability of observing *exactly* $k = [(n+1)\hat{p}]$ samples less than or equal to $x$ out of $n$ samples (where $x = F^{-1}(p)$) if the unknown parameter $P$ was indeed equal to $p$. In other words,

$$\mathsf{L}(p;\underline{I}) = P\left[M_n^{(\bar{k}+1)} \leq x \cap M_n^{(\bar{k})} > x\right] \tag{27}$$

where $M_n^{(i)}$ denotes the $i^{\text{th}}$ order statistic in a sample of size $n$ such that the maximum, $M_n = M_n^{(1)}$; and $\bar{k} = n - k = n - [(n+1)\hat{p}]$. The distribution of $M_n^{(i)}$ from a stationary dependent sequence under Conditions $D(u_n)$ and $D'(u_n)$ can be approximated as [9]:

$$P\left[M_n^{(i)} \leq u_n\right] \to e^{-\tau} \sum_{s=0}^{i-1} \frac{\tau^s}{s!} \quad \text{as } n \to \infty \tag{28}$$

$\tau$ has the same limiting form as in Eq. (7) which in this case, may be approximated as:

$$\tau \approx \theta(n-i+1)(1-p) \tag{29}$$

where, once again, $\theta$ is the extremal index. The likelihood function can be simplified using Eq. (28) as:

$$L(p;\underline{I}) = P\left[M_n^{(\bar{k}+1)} \leq x\right] - P\left[M_n^{(\bar{k}+1)} \leq x \cap M_n^{(\bar{k})} \leq x\right] = e^{-\tau}\frac{\tau^{\bar{k}}}{\bar{k}!} \tag{30}$$

which can then, using Eq. (29), be approximated, for values of $p$ close to 1, as:

$$L(p;\underline{I}) \approx c' p^{(k+1)\theta}(1-p)^{\bar{k}\theta} \tag{31}$$

where $c'$ is a constant independent of $p$.

It is known that $P$ assumes values in the interval $[0, 1]$. In the absence of any prior information on $P$, it is most logical to assume a Uniform $(0,1)$ prior distribution for $P$ [21]. Hence, the posterior density of $P$ is $f_P^{"}(p) = (c'/C) p^{(k+1)\theta}(1-p)^{\bar{k}\theta}(1)$, $0 \leq p \leq 1$, which is clearly of the Beta type, allowing us to write:

$$f_P^{"}(x;\alpha_1,\alpha_2) = \begin{cases} \dfrac{1}{B(\alpha_1,\alpha_2)} x^{\alpha_1-1}(1-x)^{\alpha_2-1} & , 0 \leq x \leq 1 \\ 0 & , \text{elsewhere} \end{cases} \tag{32}$$

The two parameters of the distribution are:

$$\alpha_1 = ([(n+1)\hat{p}]+1)\theta + 1, \quad \alpha_2 = (n-[(n+1)\hat{p}])\theta + 1 \tag{33}$$

where $n$ is the number of observations, and the estimate $\hat{p}$ is given by Eq. (25). The mean and variance of this distribution are, respectively, $\alpha_1/(\alpha_1+\alpha_2)$ and $\alpha_1\alpha_2/\{(\alpha_1+\alpha_2)^2(\alpha_1+\alpha_2+1)\}$. Hence, the updated mean of $P$ is very close to the estimate $\hat{p}$ regardless of the value of $\theta$; its variance, however, is inversely proportional to the extremal index.

In light of the above formulation, Eq. (23) can now be interpreted as the conditional distribution of the maximum of the associated i.i.d. sequence during an interval of length $t$ given fixed values of the parent distribution. Assuming the sampling uncertainty to be independent of the rate process, the unconditional distribution of the maximum of the associated i.i.d. sequence, $\hat{L}_{\max,t}$, is:

$$P\left[\hat{L}_{\max,t} \leq x\right] = \int_0^1 \int_0^\infty \exp\left[-m_t\{1-P(x)\}\right] f_{M_t}(m_t) f_P''(p) \, dm_t \, dp \qquad (34)$$

which may be estimated numerically using Monte-Carlo simulations. Finally, using the extremal index, the unconditional distribution of the maximum of the original sequence $\{L_n\}$ can be given by:

$$F_{L_{\max,t}}(x) = P\left[L_{\max,t} \leq x\right] = \left\{P\left[\hat{L}_{\max,t} \leq x\right]\right\}^\theta \qquad (35)$$

## 3. A Numerical Example of Bridge Maximum Live Load Distribution using In-Service Data

We now demonstrate the proposed methodology to estimate the maximum live load on a highway bridge using data collected under ambient traffic. The estimated maximum live load in turn can be used for (i) developing site- or region-specific design rules that may lead to more optimal designs, and (ii) conducting more accurate in-service evaluation of existing bridges. Strictly speaking, we should use the term "live load-*effect*" in the following as the collected data and the predictions both are in terms of structural strain response, we however continue with "live load" as there is no scope of confusion.

The in-service strain monitoring system used in the proposed methodology is analogous to a weigh-in-motion system, and measures peak live-load bridge strains due to site-specific traffic over extended periods of time [22, 23]. The data collection requires a minimum of equipment, no load truck, and causes no traffic restriction. The prototype system consists of a digital data acquisition system, a full-bridge strain transducer, battery pack, and an environmental enclosure. The system continuously digitizes an analog signal at 1,200 Hz, and waits for a pre-specified strain threshold to be exceeded. When this threshold is exceeded, the system evaluates the response and records the time at which the event took place, the peak strain during the event, and the area under the strain-time curve. In this research, only the peak strain during an event and its time stamp are used. A reliability-based load and resistance factor rating methodology for existing bridges involving rating at two different limit states has already been developed based on such in-service data in [24] (using a maximum live load model that did not consider correlation in the loading process).

An additional advantage of this type of data acquisition system is that it looks at structural response under ambient unrestricted traffic and thus naturally includes all possible single and multiple presence truck loading cases. The maximum live load-effect may be caused by one single heavy truck on the bridge, or the simultaneous presence of two or more trucks on the bridge:

$$L_{\max,t} = \max\left\{ L_{\max,t}^{(1)}, L_{\max,t}^{(2)}, ..., L_{\max,t}^{(m)} \right\} \tag{36}$$

where the superscript (*i*) indicates the number of trucks present simultaneously on the bridge. In the present case, in-service live-load strain data are used directly to derive statistical information on $L_{\max,t}$; hence it is not necessary to first estimate $L_{\max,t}^{(i)}, i = 1, 2, ...$ individually as would be required if the analysis started with truck weights and location on the bridge deck and went on to finding the structural response. It is thus no longer necessary here to perform additional analyses to determine the governing loading case.

The bridge selected for instrumentation and data acquisition was Bridge 1-791 which is a 3-span continuous, slab-on-steel girder structure carrying two lanes of Interstate-95 over Darley Road in Delaware. In-service strain data (peak magnitude as well as time of occurrence) was recorded at midspan of the critical girder of the approach span (beneath the right travel lane) during an approximately 11-day period in August 1998 (Figure 1a). A trigger level was set at 85 µε so that only the very large loading events would be recorded. A histogram of the data is shown in Figure 1b which represents 533 loading events.

We must acknowledge that the data set used here is clearly inadequate for practical implementation of the methodology: bridge traffic may have seasonal variations and a general upward trend over the years. We adopted this data set only as an illustrative example for the proposed methodology.

We begin by looking at the effect of raising the threshold, *u*, on the statistical properties of the loading data. Raising the threshold acts as a filter such that only those observation greater than *u*, along with their occurrence times, are retained. Statistics of the interarrival times, $\tau_i$, as a function of the threshold, *u*, starting from 85 µε up to 160 µε are analyzed to check for the asymptotic Poisson nature of the loading process. Chi-squared test of goodness of fit on the interarrival times at each threshold level saw the level of significance (at which the null hypothesis that the distribution is Exponential cannot be rejected) increase from $7 \times 10^{-5}$ at $u = 85$ to $0.05$ at $u = 160$. A preliminary analysis of the autocorrelation function of the counting processes over various intervals revealed that, (i) the dependence in the counting process falls nearly uniformly with increasing separation, although the effect of increasing threshold appears to have only a secondary effect; (ii) there appears to be a mild periodicity in each time series of around 12 hours and at integral multiples of 1, 2, 3 and 4 days; the periodicity tends to diminish with increasing threshold. We, however, do not probe this periodicity aspect further.

*3.1 The run length and the extremal index*

The extremal index of the loading sequence, $\{L_n\}$, is estimated using Eq. (13). Figure 2(a) shows the estimates, $\hat{\theta}$, at different thresholds ($u$) as a function of the run length, $r$. For any given $r$, $\hat{\theta}$ appears to approach the limit of $\theta$ as $u$ approaches $x_0$. A the same time, $\hat{\theta}$ shows a decreasing trend with increasing $r$ for any particular value of $u$, which again is consistent since the extremal index is in one sense the reciprocal of the average number of exceedances per cluster, and with increasing run length the number of cluster goes down. This points to the need of correctly identifying the run length.

We estimate the best value of the run length using Eq. (19). Three different series are used for this: (i) the peak strain series, (ii) the hourly maximum series, and (iii) the two-hourly maximum series. Note that the peak strain time series $\{L_n\}$ was recorded at random instances of time, and strictly speaking, quantities such as autocorrelation function (ACF), periodogram etc. of the loading process cannot be determined from the data. We, however, can map this series on one that is sampled at constant frequency, $\{\tilde{L}_n\}$, such that $\tilde{L}_i \equiv L_i$, $i = 1, \ldots, n$, and look at this fictitious process instead. And the *k*-hourly maxima time series is defined as the block maximum loads during successive disjoint intervals of length, $\Delta t = k$ hr:

$$L_{\max \Delta t, i} = \max \{L(\tau), (i-1)\Delta t < \tau \leq i\Delta t\} \ , \ i = 1, 2, \ldots, [t/\Delta t] + 1 \qquad (37)$$

where *t* is the total length of the observation (about 11 days in this case).

We first find the scale of fluctuation for the sequence $\{\tilde{L}_n\}$ as a function of the averaging window, *T* (Figure 2 (b)). It appears that the estimate $\hat{\tau}_c$ converges to a value of between 1 and 3 as *T* grows large. Recall that *r* is an integer equal to or greater than 2. For additional confirmation, we then look at the scale of fluctuation in the hourly maximum and two-hourly maximum loads (Figure 2 (c) and (d)). For both these sequences, the scale of fluctuation is found to converge to around 1 hour (~1×1hr and ~0.5×2hr, respectively). Now, the average number of loading events is around 2 per hour (the mean interarrival time is about 29.3 min.). Therefore, based on the above findings, we adopt a value of 2 for *r* in the following analysis.

We now obtain the minimum squared error estimate of the extremal index from the fit:

$$y(u) = \alpha + \beta_1 q(u)^{\beta_2} \qquad (38)$$

where $y(u) \equiv \hat{\theta}_R(u; r = 2)$ is obtained from Eq. (13), $\alpha \equiv \theta$ (cf. Eq. (14)), the exceedance probability $q(u) \equiv 1 - p(u)$ is estimated from Eq. (25), and $\beta_1, \beta_2$ are parameters from Eq. (16). Figure 2(e) shows the runs estimators (with $r = 2$) as a function of the exceedance probability corresponding to *u* running from 100 to 175 at increments of 5. The minimum squared error fit to the data according to Eq. (38) are also shown in the Figure, yielding a value of the extremal index as $\theta = 0.93$. This high value of θ indicates that the load sequence is almost independent (as was assumed in an earlier analysis [24]), a likely consequence of the rather high trigger of 85 microstrain set for the in-service recording device.

*3.2 Cox Process with Lognormal arrival rate*

The Cox process model for the load occurrence point process allows flexibility in tackling a large variety of load processes and at the same time is computationally efficient for generating sample functions. We assume a simple stationary model that gives non-negative smooth sample functions, namely, a lognormal process given by:

$$\Lambda(t) = \exp[\mu + \sigma z(t)] \quad (39)$$

where $\mu$ and $\sigma$ are constants, and $z(t)$ is a zero-mean unit-variance stationary Gaussian process with autocorrelation function $\rho(\tau)$. The stationary mean and coefficient of variation (c.o.v.) of $\Lambda(t)$ are, respectively, $\mu_\Lambda = \exp[\mu + \sigma^2/2]$ and $\delta_\Lambda = \sqrt{\exp(\sigma^2) - 1}$. We further assume that the autocorrelation function is exponentially decaying with correlation length, $\tau_0$, i.e., $\rho(\tau) = \exp(-|\tau|/\tau_0)$. Of course, the autocorrelation function, $\rho_\Lambda(\tau)$ of $\Lambda(t)$ is not strictly the same as that of $z(t)$: the relation between the two being given by $\rho(\tau) = \ln(1 + \rho_\Lambda(\tau)\delta_\Lambda^2)/\ln(1 + \delta_\Lambda^2)$ [25]. The difference, however, is generally small and becomes negligible for $\delta_\Lambda < 0.3$. We therefore adopt $\rho_\Lambda(\tau) \approx \rho(\tau)$ in the following. The random mean measure, $M_t$ (Eq. (20)), can then be shown to have the first two moments as:

$$E[M_t] = t\exp(\mu + \sigma^2/2) \tag{40}$$

$$\text{var}[M_t] = 2\sigma_\Lambda^2 \left\{ t\tau_0 + \tau_0^3(e^{-t/\tau_0} - 1) \right\} \tag{41}$$

where $\sigma_\Lambda^2$ is the stationary variance of the process $\Lambda$.

We now propose the following method for estimating the unknown parameters $\mu, \tau_0$ and $\sigma$ : First, estimate $E[M_t]$ for different values of $t$, minimize the error with Eq. (40) and hence obtain least square estimate for $\mu_\Lambda$. Then, estimate $\text{var}[M_t]$ for different values of $t$ and minimize the error with Eq. (41); hence obtain least square estimates for $\tau_0$ and $\sigma_\Lambda$. Use the estimates of $\sigma_\Lambda$ and $\mu_\Lambda$ in turn to obtain $\mu$ and $\sigma$.

Figure 3(a) and (b) show the estimated first two moments of $M_t$ for $t$ = 1, 2, 3, ..., 24 hrs. Non-linear least square analyses of the data yielded the following estimated parameters: $\mu_\Lambda$ = 1.99, $\sigma_\Lambda$ = 1.50 and $\tau_0$ =19.40 hrs. Thus the estimated lognormal parameters are (Eq. (39)): $\mu = 0.53$ and $\sigma = 0.56$ when $t$ is expressed in hours. Figure 3(c) shows 10 sample functions of the process $\Lambda(t)$ generated with the above parameters.

*3.3 Distribution of the maximum of the associated i.i.d. sequence*

The distribution of the maximum load, $\hat{L}_{\max,t}$, during the interval $(0,t]$ of the associated i.i.d. sequence is estimated next. Point estimates of the CDF, $\hat{p}$, of the load sequence $\{L_n\}$ (Figure 1b) at nine different strain values (*l*) are listed in Table 1. Clearly, as is desirable, only the right tail of the parent distribution has been utilized in estimating the distribution of the maximum. Further, the highest value of $l = 255$ microstrain was chosen for this exercise so as to just exceed the maximum observed response of 254.4 microstrain in the sample. A Bayesian updating of the CDF is performed (re. Eq. (32) with $\theta = 1$); the mean and the c.o.v. of the updated distribution are listed in the Table at various values of $l$. The c.o.v. of $P$ is found to become smaller and smaller as one moves along the upper tail – a result of the reasonably large sample size. We select the time interval $t = 1$ day. The unconditional CDF of the daily maximum, $\hat{L}_{\max,1d}$, of the associated i.i.d. sequence is listed in the last column of Table 1; 10,000 Monte Carlo simulations were used in estimating Eq. (34) for each value of $l$.

The maximum from an i.i.d. sequence approaches one of the three classical extreme value distributions for largest values. Of these, the Gumbel (i.e., Type I maximum) and the Frechet (i.e., Type II maximum) distributions were tried for $\hat{L}_{\max,1d}$ (Figure 4). The third, Weibull distribution for maxima, was not tried here since it is limited on the right, although this property of the Weibull distribution can be attractive in situations where geometric, bridge load posting or any other constraint can be shown to put a well-defined upper limit on the vehicular load that can be placed on the bridge. The Gumbel fit was clearly better in the present case, and was adopted for $\hat{L}_{\max,1d}$ in this paper:

$$F_{\hat{L}_{\max,1d}}(x) = \exp\left[-\exp\left(-\hat{\alpha}_{1d}(x - \hat{u}_{1d})\right)\right] \qquad (42)$$

where $\hat{\alpha}$ and $\hat{u}$ are the scale and mode, respectively, of the maximum of the associated i.i.d. sequence. Figure 4 also gives the best fit straight line through the data from which the parameters can be estimated as $\hat{\alpha}_{1d} = 0.0260$ and $\hat{u}_{1d} = 157.4$ microstrain.

*3.4 Maximum live load for various time intervals*

We are now in a position to determine the distribution of the maximum, $L_{\max,1d}$, of the actual live load sequence. Recall that the distribution of the maximum of the associated i.i.d. sequence and the actual dependent stationary sequence are of the same type (cf. Eqs. (8), (9) and (35)) differing only in terms of the shape parameter by the factor $\theta$. For the Gumbel family which is closed under maximization, this leads to an unchanged $\alpha$ (hence an unchanged variance) and a mode (and mean) shifted to the left by an amount $(1/\alpha)\ln(1/\theta)$, $0 < \theta \leq 1$. The closer $\theta$ is to 1, the more modest is this shift. Hence, the Gumbel parameters of the actual maximum, $L_{\max,rd}$, corresponding to $r$ days ($r$ = integer) can be given by:

$$\alpha_{rd} = \hat{\alpha}_{1d} \equiv \alpha$$
$$u_{rd} = \hat{u}_{1d} - \frac{1}{\alpha}\ln\left(\frac{1}{\theta}\right) + \frac{1}{\alpha}\ln(r) \tag{43}$$

The mean and c.o.v. of the maximum live load for various time intervals, $t$, up to 75 years and for various values of $\theta$ are listed in Table 2. As is well-known, with increasing $t$, the maximum live load distribution becomes narrower and shifts to the right. The case of $\theta = 1$ signifying an i.i.d. assumption, is tabulated first. The difference of the maximum load including dependence in the loading process (with its high extremal index of 0.93) from that based on an i.i.d. assumption is understandably not much in this case. For the purpose of comparison, the consequence of lower values of $\theta$, signifying increasingly greater dependence in the parent process, is also demonstrated in the Table. We can conclude that, in general, including the effect of dependence in the parent loading process decreases the mean and increases the c.o.v. of the maximum load (although the CDF is always underestimated with an i.i.d. assumption, Eq. (12)). As may be expected, the influence of dependence in the parent loading process on the maximum diminishes with increasing *t*.

## 4. Conclusions

This paper has presented a methodology that allows the use of in-service structural response data to obtain a realistic probabilistic model of extreme live loads. A considerable part of this effort has involved accounting for the possible dependence in the parent loading process. Dependence in the arrival rate process as well as in the associated load magnitudes was considered. Inclusion of such dependence in the model eliminates the unnecessary conservatism introduced by the potentially unrealistic i.i.d. assumption in the resultant maximum live load distribution. Under weak mixing type dependence, the maximum of the stationary and dependent loading process was characterized using the extremal index and the marginal distribution of the parent process. The scale of fluctuation of the loading process was used to identify clusters of exceedances above high thresholds. A Bayesian updating, derived from the distribution of order statistics in a dependent stationary series, was performed on the sample distribution function.

The methodology was used to predict maximum live load on a highway bridge. Loads (peak strain response) due to ambient traffic (including multiple presence of trucks) were monitored, and only those above a high threshold (i.e., trigger) were considered as loading events and were recorded for about 11 days. This limited data is clearly inadequate for practical implementation: we adopted this data set only as an illustrative example. The asymptotic behavior of extremes from the sample with increasing thresholds was investigated. The parameters of the random arrival rate process modeled as a Cox process with stationary lognormal intensity were determined. The effect of the extremal index, and hence of a potentially unrealistic i.i.d. assumption, on the distribution of the maximum was demonstrated. A related problem – that of predicting the asymptotic *minima* of a stationary dependent sequence using the extremal index – has recently been analyzed [26] in the context of strength of carbon nanotubes modeled with atomistic simulations.

**Tables**

**Table 1: Estimates of daily maximum peak strain**

| Right Endpoint $l$ | Marginal distribution of load | | | Distribution of max load of associated i.i.d. sequence $F_{\hat{L}\max,1\,day}(l)$ [Eq. (34)] |
|---|---|---|---|---|
| | Point Estimate $\hat{p}(l) = k/(n+1)$ [Eq. (25)] | Updated Beta Distribution parameters for $P$ [Eq. (32), $\theta = 1$] | | |
| | | mean | c.o.v | |
| 100 | 0.8202 | 0.8209 | 0.02016 | 0.00823 |
| 115 | 0.9157 | 0.9160 | 0.01306 | 0.06200 |
| 130 | 0.9476 | 0.9478 | 0.01013 | 0.14751 |
| 145 | 0.9644 | 0.9646 | 0.00827 | 0.24984 |
| 160 | 0.9738 | 0.9739 | 0.00707 | 0.35286 |
| 175 | 0.9850 | 0.9851 | 0.00531 | 0.53316 |
| 190 | 0.9888 | 0.9888 | 0.00459 | 0.61584 |
| 205 | 0.9944 | 0.9944 | 0.00324 | 0.78115 |
| 255 | 0.9981 | 0.9981 | 0.00187 | 0.92133 |

**Table 2: Maximum live load statistics for different time intervals and effect of dependence in the parent loading process**

| Time, $t$ | Statistics of maximum live load, $L_{max,t}$ | | | | | | | |
|---|---|---|---|---|---|---|---|---|
| | $\theta = 1$ | | $\theta = 0.93$ | | $\theta = 0.75$ | | $\theta = 0.50$ | |
| | mean | c.o.v. | **mean** | **c.o.v.** | mean | c.o.v. | mean | c.o.v. |
| 1 day | 178.0 | 0.275 | **177.0** | **0.279** | 168.7 | 0.293 | 153.3 | 0.322 |
| 1 year | 407.7 | 0.121 | **404.1** | **0.122** | 395.9 | 0.125 | 380.1 | 0.130 |
| 2 years | 434.5 | 0.114 | **430.8** | **0.115** | 422.6 | 0.117 | 406.7 | 0.121 |
| 10 years | 496.6 | 0.0997 | **492.8** | **0.100** | 484.6 | 0.102 | 468.6 | 0.105 |
| 50 years | 558.7 | 0.0886 | **554.7** | **0.0890** | 546.6 | 0.0904 | 530.5 | 0.0930 |
| 75 years | 574.4 | 0.0862 | **570.3** | **0.0866** | 562.2 | 0.0879 | 546.1 | 0.0903 |

**Figure captions**

Figure 1: (a) Time-line of loading events spanning 11 days in August 1998 on Bridge 1-791 (b) peak strain histogram and (c) interarrival times of the 533 loading events

Figure 2: Estimating the extremal index, $\theta$: (a) importance of run length, r, at different thresholds (u, microstrain) in estimating $\theta$; (b –d ) Estimating r from the scale of fluctuation of three sequences; (e) Minimum squared error estimate of $\theta$ ( for r = 2)

Figure 3: Estimates of (a) mean and (b) variance of the random mean measure, $M_t$, of the load arrival Cox process (circle = data, line = least square fit), and (c) ten sample functions of the Cox process $\Lambda(t)$ based on the parameters estimated from (a) and (b)

Figure 4: (a) Gumbel and (b) Frechet probability fit of daily maximum of the associated i.i.d. live-load sequence